\documentclass[12pt]{article}
\usepackage{epsfig, amssymb}
\usepackage{graphicx,epsfig}
\usepackage{color} 
\begin{document}
\thispagestyle{empty}
\begin{flushright}
\end{flushright}

\bigskip

\begin{center}
\noindent{\Large \textbf
{A conformal scalar $n$-point function in momentum space}}\\ 
\vspace{2cm} \noindent{
Jae-Hyuk Oh\footnote{e-mail:jack.jaehyuk.oh@gmail.com}}

\vspace{1cm}
  {\it
Department of Physics, Hanyang University, Seoul 133-791, Korea\\
 }
\end{center}

\vspace{0.3cm}
\begin{abstract}
We suggest a certain type of conformal $n$-point function of scalar primaries where the scalar operators share the same scaling dimension. The conformal correlation functions are obtained in momentum space, and we show that they satisfy the conformal Ward identities.
\end{abstract}


\paragraph{Introduction and motivation}
There are some of literatures appearing these days studying conformal correlation functions in momentum space\cite{Bzowski:2013sza,Bzowski:2019kwd,Maglio:2019grh,Oh:2013tsa,Oh:2015xva}. The motivation for such a study may be two-folds. First, if one analyzes the correlation functions in momentum space, it may be possible to interpret them in Feynman-like diagrammatic ways. For each vertex point, momentum conservation must be hold and so one can see the flows of momenta to inward and(or) outward of the vertex. The second motivation is holography. In many of the holographic computations, they perform Fourier transform from position space to momentum space in conformal boundary of the AdS space. 
In many cases the holographic correlation function contains differential operators acting on the boundary fields in it. Sometimes, they are non-local, showing the differential operators in the denominator of the correlation function. In such cases, the expression may be more clear in momentum space.

In this note, we claim that in $d$-dimensional Euclidean space, for a scalar operator $O_{\Delta}(p)$ whose conformal dimension is $\Delta=\frac{d+1}{2}$, the following object becomes a $n$-point correlation function of those operators:
\begin{equation}
\langle O_{\Delta}(p_1)O_{\Delta}(p_2)...O_{\Delta}(p_{n-1})O_{\Delta}(p_n) \rangle=\frac{1}{\left(
\sum_{i=1}^{n-1}|p_i|+\left| \sum_{j=1}^{n-1} p_j \right|
\right)^{(n-1)d-\frac{n}{2}(d+1)}},
\end{equation}
where $p_n=-\sum_{j=1}^{n-1}p_j$ for momentum conservation. In the rest of this note, we prove this claim.

\paragraph{A momentum space correlators of scalar primaries }
Now, we suggest a form of the conformal correlation function of scalar primaries which share the same scaling dimension, $\Delta$. We consider an object being given by
\begin{equation}
\label{correlation}
\Phi(p_1,p_2,...,p_{n-1};p_n)=\frac{1}{\left(
\sum_{i=1}^{n-1}|p_i|+\left| \sum_{j=1}^{n-1} p_j \right|
\right)^\alpha},
\end{equation}
where the $|p_i|\equiv \sqrt{p_{i\eta} p_{i\gamma}
\delta^{\eta\gamma}}$(the index $i$ is not summed) is the absolute value of the Euclidean momentum $p_{i\eta}$, the Greek super(sub)scripts represent the spatial indices of the momenta $p_{i\eta}$ and $\delta_{\eta\gamma}=\delta^{\eta\gamma}=\delta^\eta_\gamma$.
The $\alpha$ in the exponent is a real number and the index $i$ is for labeling of the $n$-different momenta $p_i$. Again, $p_n=-\sum_{i=1}^{n-1}p_i$, so we have $n-1$ independent momenta in it due to momentum conservation.

Since the expression is a function of the absolute values of the Euclidean momenta, it is translation and rotational invariant. Now, we check if this object transforms appropriately under dilatation and special conformal transformations. Dilation Ward identity is given by
\begin{equation}
\mathcal D \Phi(p_1,p_2,...,p_{n-1};p_n)=0,
\end{equation}
where
\begin{equation}
\mathcal D=\sum_{j=1}^{n-1}p^\eta_j \frac{\partial}{\partial p^\eta_j}+\bar\Delta,
\end{equation}
where the Greek indices are contracted by the Kroneker's $\delta$ that we introduce above.
The $\bar\Delta$ is
\begin{equation}
\bar\Delta=-\sum_{i=1}^{n}\Delta_i+(n-1)d.
\end{equation}
We are interested in a case that all the $\Delta_i=\Delta$ are the same. When one applies the operator $\mathcal D$ to the $\Phi(p_1,p_2,...,p_{n-1};p_n)$, then one gets
\begin{equation}
\label{D-eq}
\mathcal D \Phi(p_1,p_2,...,p_{n-1};p_n)=\frac{\{-\alpha-n\Delta+(n-1)d\}}{u^{\alpha+1}},
\end{equation}
where
\begin{equation}
u=
\sum_{i=1}^{n-1}|p_i|+\left| \sum_{j=1}^{n-1} p_j \right|.
\end{equation}
The last step is to check if the $\Phi(p_1,p_2,...,p_{n-1};p_n)$ satisfies special conformal Ward identity. The special conformal Ward identity in momentum space is given by
\begin{equation}
\mathcal K^k \Phi(p_1,p_2,...,p_{n-1};p_n)=0,
\end{equation}
where
\begin{equation}
\mathcal K^\kappa=\sum_{j=1}^{n-1}\left\{ 
2(\Delta_j-d)\frac{\partial}{\partial p_j^k} + p_j^\kappa\frac{\partial^2}{\partial p_j^\eta \partial p_j^\eta}-2p_j^\eta\frac{\partial^2}{\partial p_j^\eta \partial p_j^\kappa}
\right\}
\end{equation}
and we request $\Delta_j=\Delta$ for all the $j$-index. Application of the operator, $\mathcal K^\kappa$ to the $\Phi(p_1,p_2,...,p_{n-1};p_n)$ results in
\begin{eqnarray}
\label{K-eq}
\mathcal K^k \Phi(p_1,p_2,...,p_{n-1};p_n)=\frac{1}{u^{\alpha+1}}
\sum_{j=1}^{n-1}\frac{p_j^\kappa}{|p_j|}(d-2\Delta+1)\alpha
 \\ \nonumber
+\frac{1}{u^{\alpha+1}}\frac{\sum_{j=1}^{n-1}p_j^\kappa}{\left|\sum_{j=1}^{n-1}p_j\right|}
\left\{2(d-\Delta)(n-1)-d+1-2(\alpha+1)\right\}\alpha.
\end{eqnarray}

Therefore, we have three algebraic conditions from (\ref{D-eq}) and (\ref{K-eq}), under which the $\Phi(p_1,p_2,...,p_{n-1};p_n)$ becomes a conformal correlation function. The conditions that the $\Phi(p_1,p_2,...,p_{n-1};p_n)$ needs to satisfy are
\begin{itemize}
\item $-\alpha-n\Delta+(n-1)d=0$,
\item $(d-2\Delta+1)\alpha=0$,
\item $\left\{2(d-\Delta)(n-1)-d+1-2(\alpha+1)\right\}\alpha=0$.
\end{itemize}

\paragraph{Possible solution}
The three algebraic equations are not independent one another. In fact, the two of them are linearly independent. The possible solution is
\begin{itemize}
\item For a given the spatial dimension $d$, one can consider a scalar opertor $O_{\Delta}(p)$ with $\Delta=\frac{d+1}{2}$. A possible $n$-point correlation function of the operators is a form of (\ref{correlation}) with $\alpha=(n-1)d-\frac{n(d+1)}{2}$.
\end{itemize}

\paragraph{Examples}
\begin{itemize}
\item Holographic computations of two point function of  scalar operators by employing conformally coupled scalar in AdS space as a gravity dual is given by
\begin{equation}
\langle O_{\Delta}(p_1)O_{\Delta}(-p_1) \rangle=C_2|p_1|,
\end{equation}
where the scaling dimension of the operators is $\Delta=\frac{d+1}{2}$ for a constant $C_2$\cite{Oh:2014nfa,Oh:2012bx,Jatkar:2013uga}. This also can be obtained from the general expression of the conformal two point function in momentum space given in \cite{Bzowski:2013sza}.

\item In $d=3$, for a scalar operator with scaling dimension $\Delta=2$, holographic computations of four point function by using conformally coupled scalar in AdS$_4$ as a gravity dual has a form of 
\begin{equation}
\langle O_{\Delta}(p_1)O_{\Delta}(p_2)O_{\Delta}(p_{3})O_{\Delta}(-p_1-p_2-p_3) \rangle=\frac{C_4}{\left(
\sum_{i=1}^{3}|p_i|+\left| \sum_{j=1}^{3} p_j \right|
\right)},
\end{equation}
upto some constant $C_4$\cite{Oh:2014nfa}.

\item In $d=5$, for a scalar operator with scaling dimension $\Delta=3$, holographic computations of three point function obtained from its a gravity dual, conformally coupled scalar in AdS$_6$  has a form of 
\begin{equation}
\langle O_{\Delta}(p_1)O_{\Delta}(p_2)O_{\Delta}(p_{3})\rangle=\frac{C_3}{\left(
\sum_{i=1}^{2}|p_i|+\left| \sum_{j=1}^{2} p_j \right|
\right)},
\end{equation}
upto some constant $C_3$\cite{ohappear}. This also can be obtained from the general expression of the conformal three point function(so called K-triple integral) in momentum space given in \cite{Bzowski:2013sza}.
\end{itemize}


\section*{Acknowledgement}
J.H.O thanks his $\mathcal W.J$ and $\mathcal Y.J$.  This work was supported by the National Research Foundation of Korea(NRF) grant funded by the Korea government(MSIP) (No.2016R1C1B1010107) and Research Institute for Natural Sciences, Hanyang University.

\end{document}